\documentclass[aps,prd,twocolumn,showpacs,amsmath,amssymb]{revtex4-1}
\usepackage{graphicx}
\usepackage{color}

\newcommand{\LQCD}{\Lambda_{\text{QCD}}}
\newcommand{\calCP}{\mathcal{CP}}
\newcommand{\calP}{\mathcal{P}}
\newcommand{\calC}{\mathcal{C}}

\newcommand{\bsigma}{\boldsymbol{\sigma}}

\newcommand{\bj}{\boldsymbol{j}}
\newcommand{\bA}{\boldsymbol{A}}
\newcommand{\bB}{\boldsymbol{B}}
\newcommand{\bE}{\boldsymbol{E}}
\newcommand{\bp}{\boldsymbol{p}}

\newcommand{\bq}{\boldsymbol{q}}
\newcommand{\bx}{\boldsymbol{x}}
\newcommand{\boldj}{\boldsymbol{j}}
\newcommand{\feyn}[1]{
  \setbox0=\hbox{\ensuremath{#1}}
  \hbox to\wd0{\hbox to0pt{\hbox to\wd0{\hss/\hss}\hss}\box0}}
\begin{document}

\title{What flows in the chiral magnetic effect?\\ --- Simulating the
  particle production with $\calCP$-breaking backgrounds ---}

\author{Kenji Fukushima}
\affiliation{Department of Physics, The University of Tokyo,
             7-3-1 Hongo, Bunkyo-ku, Tokyo 113-0033, Japan}

\begin{abstract}
 To address a question of whether the chiral magnetic current is a
 static polarization or a genuine flow of charged particles, we
 elucidate the numerical formulation to simulate the net production of
 right-handed particles and anomalous currents with $\calCP$-breaking
 background fields which cause an imbalance between particles and
 anti-particles.  For a concrete demonstration we numerically impose
 pulsed electric and magnetic fields to confirm our answer to the
 question that the produced net particles flow in the dynamical chiral
 magnetic effect.  The rate for the particle production and the chiral
 magnetic current generation is quantitatively consistent with the
 axial anomaly, while they appear with a finite response time.  We
 emphasize the importance to quantify the response time that would
 suppress observable effects of the anomalous current.
\end{abstract}
%\pacs{}
\maketitle

%%%%%%%%%%   Introduction   %%%%%%%%%%
\section{Introduction}

In many quantum problems in physics it is highly demanded to establish
a numerical framework to simulate full dynamical processes of particle
production out of equilibrium.  Unsettled and important problems
include the leptogenesis and the baryogenesis for the explanation of
the baryon asymmetry of the Universe (BAU).  It is well-known that
Sakharov's three conditions are needed for the BAU; namely,
$B$-breaking process, $\calC$ and $\calCP$ violation, and
out-of-equilibrium.  The particle production on top of the electroweak
sphaleron~\cite{Klinkhamer:1984di} accommodates a process with
$\Delta(B+L)\neq 0$, which means that $B+L$ decays toward
chemical-equilibrium [and this is why the idea based on the SU(5)
grand unified theory~\cite{Yoshimura:1978ex} does not work out for the
BAU].  Therefore, if the leptogenesis with some $(B-L)$-breaking
process beyond the Standard Model (see Ref.~\cite{Fukugita:1986hr} for
example) generates $L\neq 0$ initially, it would amount to $B\neq 0$
finally in chemical-equilibrium.  The particle production associated
with the sphaleron transition has a character of topological
invariance and the index theorem relates the topological number to the
change in the particle number.  We must recall, however, that the
sphaleron is a special static configuration at saddle-point and for
more general and non-topological gauge configurations with $\calC$-
and $\calCP$-odd components we have to consider microscopic
simulations with Weyl fermions.  Also, to compute the momentum spectrum
of produced particles with arbitrary gauge backgrounds, there is no
longer mathematical elegance and some brute-force tactics with
numerical simulations is indispensable.

The sphaleron transition rate is proportional to $T^4$ where $T$ is
the temperature~\cite{Arnold:1987zg}, and so such topological
fluctuations should be abundant also in the strong interaction when
the physical system is heated up to $T>\LQCD$ as argued in
Ref.~\cite{McLerran:1990de}.  Such high temperature is indeed realized
in the quark-gluon plasma created in the relativistic heavy-ion
collision (HIC) experiment.  Although the strong interaction does not
break $\calCP$ (except for negligibly small strong $\theta$), we can
still anticipate local violation of $\calP$ and $\calCP$ before
thermalization~\cite{Kharzeev:1998kz}.  This possibility is nowadays
referred to commonly as the local parity violation (LPV).  Theoretical
and experimental challenges are still ongoing about the LPV
detection (and no direct evidence has been found for it).  Along these
lines the recognition of the interplay with pulsed magnetic field
$\bB$ between two heavy ions was a major
breakthrough~\cite{Kharzeev:2007jp};  $\bB$ on top of $\calCP$-odd
backgrounds would induce an electric current $\boldj$ in parallel to
$\bB$, which is summarized in a compact formula of the chiral magnetic
effect (CME) with chiral chemical potential
$\mu_5$~\cite{Fukushima:2008xe}.  In a realistic situation of the HIC
we can anticipate $\calP$ and $\calCP$ violation not only from the QCD
sphalerons but also from the glasma~\cite{Kharzeev:2001ev} which is a
description of the initial condition of the HIC in terms of coherent
fields.  Then, without introducing $\mu_5$ that has huge theoretical
uncertainty (see Ref.\cite{Hirono:2014oda} for a recent study with
$\mu_5$), we can and should directly compute the anomalous currents
associated with the particle production with $\calCP$-odd backgrounds
composed from chromo-electric and chromo-magnetic
fields~\cite{Fukushima:2010vw}.

It would be an intriguing attempt to test the CME in a table-top
experiment in a way similar to the quantum Hall effect.  The biggest
difference between the CME and the Hall effect arises from the fact
that the carriers of electric charge are not the ordinary particles
but chiral particles;  that is, a non-zero excess of the right-handed
particle number to the left-handed particle number is
required~\cite{Kharzeev:2013}.  It is quite non-trivial how to
implement such a chiral battery (usually represented by $\mu_5\neq 0$)
in materials.  Just recently the CME may have been confirmed in a
system rather similar to the glasma, in which chirality imbalance is
imposed by background fields that break $\calCP$
symmetry~\cite{Li:2014}.

Here, to elucidate the motivation of our present study clearly, let us
discuss theoretical subtleties about the interpretation of the chiral
magnetic (and separation) effect.
\begin{enumerate}
\item What carries the electric charge of the current?  In the
  derivation using the Chern-Simons-Maxwell
  theory~\cite{Kharzeev:2009fn} the chiral magnetic current appears in
  Amp\`{e}re's law, but it takes a similar form as Maxwell's
  displacement current.  We know that Maxwell's displacement current
  is a source for the magnetic field but no charge carrier flows
  (though the Poynting vector shows a flow of electromagnetic energy)
  as pointed out in Ref.~\cite{Fukushima:2012vr}.

\item What is the momentum spectrum of charged particles that flow as
  the chiral magnetic current?  In the derivation using the
  thermodynamic potential~\cite{Fukushima:2008xe} a finite result
  remains after a cancellation between infinitely large momentum
  contributions $p^z\sim\pm\infty$, but it is unlikely in any
  experiment that particles with such large momenta emerge from the
  vacuum.  In the kinetic
  derivation~\cite{Son:2012wh,Stephanov:2012ki,Gao:2012ix} a thermal
  distribution of particles is finally assumed to retrieve the exact
  CME coefficient, which provides us with a theoretically correct
  description of the dynamical current.  It is, however, still needed
  to understand how such a distribution of particles arises.  In
  general with arbitrary (non-thermal) initial conditions the particle
  distribution functions are not necessarily thermal and then the CME
  current may not obey the standard formula, which also motivates
  the numerical simulation.

\item The expression $\langle\Omega|\bj|\Omega\rangle\neq0$ does not
  always mean a flowing current but it may represent a polarization
  that is a static expectation value of a vector operator $\bj$, where
  $|\Omega\rangle$ represents a certain pure state of the physical
  system.  (We can generalize our discussions to mixed states, but the
  consideration with a pure state suffices for our purpose.)  In the
  lattice-QCD simulation in Euclid spacetime as in
  Ref.~\cite{Yamamoto:2011gk}, $|\Omega\rangle$ is the Euclidean QCD
  vacuum, and the current measured with $|\Omega\rangle$ should be a
  polarization because the current is a non-equilibrium and steady
  dynamical phenomenon.  Actually the chiral separation effect
  $\langle\Omega|\bj_{\rm A}|\Omega\rangle\propto \mu\bB$ has a
  natural interpretation as a spin polarization~\cite{yamamoto} (see
  also Ref.~\cite{Yamamoto:2015fxa} for a recent study on this).
  To make a possible difference explicit we point out an example found
  in the computation of the chiral magnetic conductivity
  $\sigma_\chi(\omega,\bp)$ that represents a linear response in the
  presence of spacetime dependent $\bB(\omega,\bp)$.  From Fig.~2 of
  Ref.~\cite{Kharzeev:2009pj} it is obvious that the static limit
  $\sigma_\chi(\omega=0,\bp\to0)$ takes a value different from the
  dynamical limit $\sigma_\chi(\omega\to0,\bp=0)$.

\item How can the particle motion be aligned to the magnetic field?
  The conventional explanation (i.e., the correlation between the
  helicity and the spin alignment under a strong $\bB$) is based on a
  static picture corresponding to the polarization phenomenon.  If
  there are free right-handed fermions, say, in the direction
  perpendicular to $\bB$, they just move on a circle with the Larmor
  radius classically.  The spin receives a torque from the Berry's
  curvature term, but the spin cannot be aligned to $\bB$ without
  dissipation.  Also we make a comment that the kinetic derivation may
  look like a problem of one-particle motion, but it should be
  justified by the worldline formalism on the quantum level, and then
  the (proper) time may be given a meaning different from the genuine
  time.

\item What is the response time for the chiral magnetic current to get
  activated?  It is unlikely that the current suddenly starts flowing
  as soon as $\bB$ and $\mu_5$ are turned on.  (This is actually an
  unavoidable problem to simulate the chiral magnetic effect assuming
  a certain $\mu_5$.)  The current generation rate has been calculated
  only in an idealized setup, and the quantitative estimate of this
  response time should be crucial for experimental detection in
  realistic and thus disturbed environments.
\end{enumerate}

%---   figure   ---%
\begin{figure}
 \includegraphics[width=0.6\columnwidth]{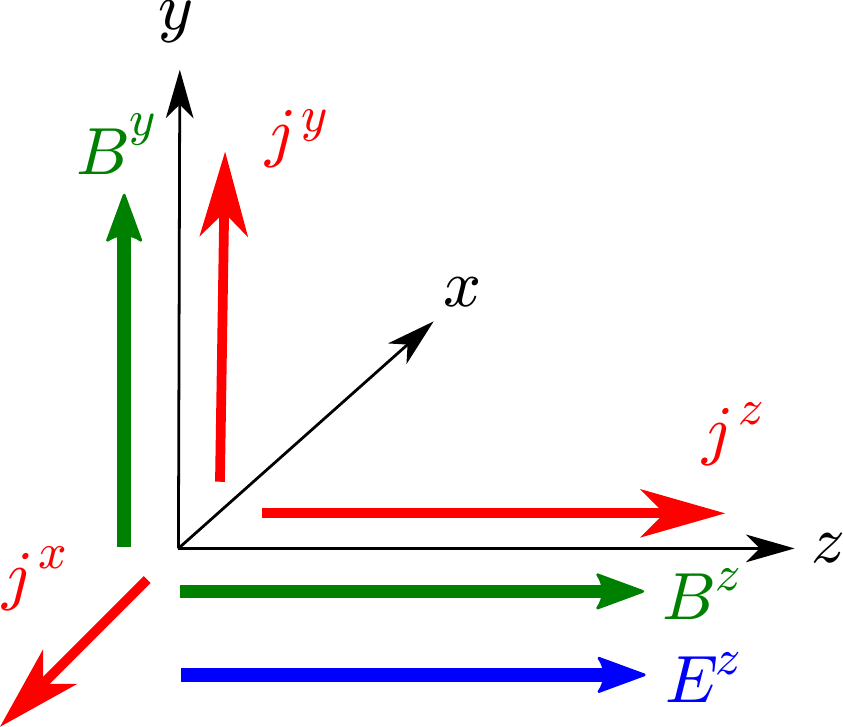}
 \caption{Schematic view of currents induced by $B^y$ in a
   $\calCP$-odd domain realized by parallel $E^z$ and $B^z$.  The
   currents flow in all the $x$, $y$, and $z$ directions;  $j^x$
   is the anomalous Hall current and $j^y$ is the CME current.}
 \label{fig:schematic}
\end{figure}
%---   figure   ---%

To answer these question, in this work, we consider a special setup as
illustrated in Fig.~\ref{fig:schematic}, which is the schematic view
of the CME setup in the HIC and in the condensed matter experiment.
The parallel $\bE$ and $\bB$ (in the $z$ direction in
Fig.~\ref{fig:schematic}) form $\calP$- and $\calCP$-odd product
$\bE\cdot\bB$, and the CME current $j^y$ is induced in a direction
perpendicular to $\bE$, which is reminiscent of the anomalous Hall
current $j^x$.  (In this sense one can regard the CME as a 3D
extension of the Hall effect.)  In the HIC $E^z$ and $B^z$ are Abelian
projected chromo-electric and chromo-magnetic
fields~\cite{Fukushima:2010vw}.  In the sphaleron transition these
fields are provided from the Abelian projected part of the sphaleron
gauge configuration too.

Then, we can address the questions in a very concrete way and we shall
present short answers in order:
\begin{enumerate}
\item The produced particles from the vacuum fluctuations form a
  current.  The right-handed particles and anti-particles have
  asymmetric momentum distributions due to $\bE\cdot\bB\neq0$, where
  the particle number is conserved with the opposite excess from the
  left-handed sector (and the current is doubled).  Therefore, the
  chiral magnetic effect can be realized in a non-equilibrium
  environment and is most relevant to the early-time dynamics of the
  HIC involving the particle production.

\item Produced particles are accelerated by $\bE$ and the momentum
  should have a peak around an impulse given by the product of the
  Lorentz force times the duration, which is the ordinary situation in
  the Schwinger mechanism with pure $\bE$.  In the same way we can
  compute the distribution function with $\bB\neq0$ as well as
  $\bE\neq0$ that is localized in momentum space and so regions with
  infinitely large momenta are completely irrelevant.

\item The particle production process is the real-time phenomenon out
  of equilibrium.  In this sense, according to our understanding based
  on the particle production,
  $\langle\Omega|\bj_A|\Omega\rangle \propto \mu\bB$ can be
  interpreted as a dynamical flow of the chirality, which results
  simply from a different combination of the right-handed and the
  left-handed sectors of the present calculation.  This fact supports
  the realization of non-zero quadrupole moment of the electric charge
  suggested from the chiral magnetic wave.  We note that if
  $\langle|\Omega|\bj_A|\Omega\rangle$ were a spin polarization or a
  magnetization, there would be no movement of chirality in space,
  which is not the case in our fully dynamical setup.

\item The particle motion has a peculiar angle distribution or
  asymmetric momentum distribution (which will be demonstrated in
  Fig.~\ref{fig:dist}) reflecting the configurations of $\bE$ and
  $\bB$.  Thus, unlike the conventional explanation, there is no need
  to bend the motion of classical particles nor the direction of the
  spin polarization with $\bB$ only.

\item One might think that the asymmetric momentum distribution occurs
  from the beginning of the particle production and so there is no
  delay for the topological currents to flow.  The plane wave,
  however, should be deformed into the Landau wave-function after a
  finite $\bB$ is switched on.  The response time is governed by the
  development in the amplitude of the Landau wave-function.  In
  particular the onsets of the particle production and the current
  generation may look different due to different rates.
\end{enumerate}

%%%%%%%%%%   Theory of the particle and current production   %%%%%%%%%%
\section{Theory of the particle and current production}

In this work we focus on the right-handed sector only and the current
(net particle) density should be doubled (canceled) once the
left-handed sector is taken into account.  The right-handed Weyl
fermion should satisfy the following equation of motion:
\begin{equation}
 \bigl( i\sigma^\mu\partial_\mu - e\sigma^\mu A_\mu \bigr)
  \phi_{\rm R} = 0
\label{eq:Weyl_eq}
\end{equation}
with $\sigma^\mu=(1,\bsigma)$.  We can readily construct a complete
set of plane-wave solutions of Eq.~\eqref{eq:Weyl_eq} for the
asymptotic states where the interaction is turned off.  In general a
constant background $A_\mu$ may be coupled even without interaction.
In the present work we set a gauge that makes $A_0=0$ for a technical
reason.  We can then write positive-energy particle solutions as
$\phi_{\rm R}=u_{\rm R}(\bp;\bA) e^{-ip\cdot x}$ with
\begin{equation}
 u_{\rm R}(\bp;\bA) = \begin{pmatrix}
  \sqrt{|\bp_A|+p_A^z} \\
 e^{i\theta(\bp_A)}\sqrt{|\bp_A|-p_A^z} \end{pmatrix}
\label{eq:ur}
\end{equation}
in a certain gauge [or a convention for the overall U(1) phase].  Here
we defined $\bp_{\pm A}\equiv \bp\mp e\bA$ and the phase factor is:
\begin{equation}
 e^{i\theta(\bp_A)}
  = \frac{p_A^x + ip_A^y}{\sqrt{(p_A^x)^2 + (p_A^y)^2}} \;.
\label{eq:phase}
\end{equation}
We can identify the anti-particle state from
$\phi_{\rm\bar{R}}=-i\sigma^2\phi_{\rm R}^\ast$, which leads to a
relation: $u_{\rm \bar{R}}(\bp;\bA)=u_{\rm R}(-\bp;\bA)$.  In the same
way we can find the negative-energy particle and anti-particle
solutions with
$\phi_{\rm R/\bar{R}}=v_{\rm R/\bar{R}}(\bp;\bA) e^{+ip\cdot x}$,
which results in
$v_{\rm R}(\bp;\bA)=-e^{-i\theta(\bp_{-A})}u_{\rm R}(\bp;-\bA)$ and
$v_{\rm\bar{R}}(\bp;\bA)=-e^{-i\theta(\bp_A)} u_{\rm R}(-\bp;-\bA)$.
We note that $u_{\rm R}(\bp;\bA)$ and $v_{\rm \bar{R}}(\bp;\bA)$ have
an energy $\pm |\bp_A|$, while other two, $u_{\rm \bar{R}}(\bp;\bA)$
and $v_{\rm R}(\bp;\bA)$, have an energy $\pm |\bp_{-A}|$.

For the problem of particle and anti-particle production we evaluate
an amplitude for the transition from a negative-energy state (with
momentum $\bp$ and vector potential $\bA$) to a positive-energy state
(with momentum $\bq$ and vector potential $\bA'$), which we can
explicitly express as
\begin{equation}
 \begin{split}
 & \beta_{\bq,\bp} = \int d^3\bx\,
  \frac{u_{\rm R}^\dag(\bq_{A'}) e^{i|\bq_{A'}|x^0 + i\bq\cdot\bx}}
  {\sqrt{2|\bq_{A'}|}} \, g_{-\bp}(x^0,\bx) \;,\\
 & \bar{\beta}_{\bq,\bp} = \int d^3\bx\,
  \frac{u_{\rm \bar{R}}^\dag(\bq_{-A'}) e^{i|\bq_{-A'}|x^0 + i\bq\cdot\bx}}
  {\sqrt{2|\bq_{-A'}|}} \, \bar{g}_{-\bp}(x^0,\bx)
 \end{split}
\end{equation}
for the particles and the anti-particles, respectively.  We here
introduced new functions, $g_{\bp}(x)$ and $\bar{g}_{\bp}(x)$, as
solutions of the particle and the anti-particle equations of motion
satisfying the following negative energy boundary conditions:
\begin{equation}
 \begin{split}
 & g_{\bp}(x) \:\longrightarrow\:
  \frac{v_{\rm R}(\bp_{-A}) e^{i|\bp_{-A}|x^0 - i\bp\cdot\bx}}
  {\sqrt{2|\bp_{-A}|}}\;,\\
 & \bar{g}_{\bp}(x) \:\longrightarrow\:
  \frac{v_{\rm \bar{R}}(\bp_A) e^{i|\bp_A|x^0 - i\bp\cdot\bx}}
  {\sqrt{2|\bp_A|}}
 \end{split}
\end{equation}
for $x^0$ around the initial time in the past.  It should be noted
that the equation of motion for the anti-particle is not
Eq.~\eqref{eq:Weyl_eq} but $e$ is replaced with $-e$ and $\sigma^\mu$
with $\bar{\sigma}^\mu$.  Finally we can express the net particle
number (i.e.\ the particle number minus the anti-particle number
\textit{in the right-handed sector}) as well as the spatial currents
in the following manner (after dropping the zero-point oscillation
term from $J^0$):
\begin{equation}
 \begin{split}
 & J^0 = e \int\frac{d^3\bq}{(2\pi)^3} \bigl[ f(\bq)
  - \bar{f}(\bq) \bigr] \;,\\
 &\boldsymbol{J} = e \int\frac{d^3\bq}{(2\pi)^3} \biggl[
  \frac{\bq_{A'}}{|\bq_{A'}|} f(\bq)
  - \frac{\bq_{-A'}}{|\bq_{-A'}|} \bar{f}(\bq) \biggr]\;,
 \end{split}
\label{eq:currents}
\end{equation}
where the distribution functions $f(\bq)$ and $\bar{f}(\bq)$ above are
defined with the amplitudes integrated over all incoming momenta,
i.e.,
\begin{equation}
 f(\bq) \equiv \int\frac{d^3\bp}{(2\pi)^3}\,|\beta_{\bq,\bp}|^2
  ,\quad\!
 \bar{f}(\bq) \equiv \int\frac{d^3\bp}{(2\pi)^3}\,
  |\bar{\beta}_{\bq,\bp}|^2 \;.
\end{equation}
These functions are naturally given an interpretation as the
distribution functions of particles and anti-particles.  We note that
$J^0$ above is a formula equivalent to that as utilized in
Ref.~\cite{Gelis:2004jp}.  We also make a comment on the relation to
the chiral kinetic theory where the current is defined with
$\dot{\bx}$ and the equation of motion solves $\dot{bx}$ in terms of
$\bq$ and an additional contribution from Berry's curvature.  The
appearance of the \textit{canonical} momenta $\bq_{A'}$ (shifted by
$\bA'$) in Eq.~\eqref{eq:currents} corresponds to using $\dot{bx}$ on
the level of the equation of motion.  Still, it is highly non-trivial
how to see the analytical relation between two approaches, which will
be an interesting future problem.

%%%%%%%%%%   Simulation setup   %%%%%%%%%%
\section{Simulation setup}

We here consider pulsed fields approximated by step functions for a
duration $T$.  We choose the origin of the time so that we start
solving Eq.~\eqref{eq:Weyl_eq} numerically from $t=0$.  Introducing a
temporal profile defined by
\begin{equation}
 \epsilon(t) \equiv \begin{cases} 1 & (-T/2 \,\leq\, t-t_0 \,\leq\, T/2) \\
                             0 & (|t-t_0| > T/2) \;,
 \end{cases}
\end{equation}
we can explicitly specify the electric and magnetic fields that we
want to consider here as
\begin{equation}
 E^z(t) \sim B^z(t) \sim B^y(t) \propto \epsilon(t)\;.
\label{eq:exfield}
\end{equation}
This choice of the temporal profile is motivated based on the glasma
dynamics in which both the external magnetic field and the
chromo-fields decay within the same time scale of
$\sim 0.1\;\text{fm}/c$~\cite{Lappi:2006fp,Fukushima:2007yk}.
Theoretically speaking, to define the produced particle number
uniquely, we should setup the asymptotic states where interaction is
switched off.  We also mention another possibility to prescribe the
particle number in a transient state (mostly relying on an adiabatic
approximation) as studied in Ref.~\cite{Tanji:2008ku}, though we do
not adopt it.

It is, however, technically non-trivial how to realize the external
fields as Eq.~\eqref{eq:exfield} strictly.  Here, we will make a
simpler choice of the gauge potentials:
\begin{equation}
 \begin{split}
 & A^x = B_\perp \epsilon(t) z - B_\parallel \epsilon(t) y\;,\quad
   A^y = 0\;,\\
 & A^z = -E_0 \bar{\epsilon}(t)\;,
 \end{split}
\label{eq:vector}
\end{equation}
where $\bar{\epsilon}(t)=\int_{-\infty}^t \epsilon(t')dt'$.  We note
that $A^z(t=-\infty)=0$ and $A^z(t=\infty)=-E_0 T\neq 0$ and this is
why we should keep $\bp_A$ in all the expressions for produced
particles.  The above vector potentials actually lead to the magnetic
fields as almost desired:
\begin{equation}
 B^x(t) = 0\;,\quad
 B^y(t) = B_\perp \epsilon(t)\;,\quad
 B^z(t) = B_\parallel \epsilon(t)\;,
\end{equation}
whereas the electric fields have some extra components:
\begin{equation}
 \begin{split}
 & E^x(t) = -B_\perp \epsilon'(t) z + B_\parallel \epsilon'(t) y\;,\quad
   E^y(t) = 0\;,\\
 & E^z(t) = E_0 \epsilon(t)\;.
 \end{split}
\end{equation}
We chose the gauge~\eqref{eq:vector} so that we can have $E^y(t)=0$
because we are interested in $j^y$.  This also indicates that the
current $j^x$ has not only the anomalous component but a finite
contribution induced by $E^x\neq0$.

We note that in our present calculation we neglect the back-reaction
and so this $j^x$ would not affect $j^y$.  Because we start the
simulation with the null initial condition with zero particle, this
approximation for the back-reaction is expected to hold well
especially for our setup with pulsed fields.  In principle our
formulation can be upgraded to the classical statistical simulation
containing back-reaction~\cite{Hebenstreit:2013baa} or the Dirac
equations can be combined with a numerical solution of Maxwell's
equations.  The classical evolution suffices for our present purpose
to figure out the net particle production and associated CME current
generation.

Let us now go into some more technical parts.  It is convenient to
keep the reflection symmetry of axes in order to avoid non-vanishing
currents due to lattice artifact.  A perturbation theory in terms of
$A^i$ would give a term like $\propto z, y$ that should be vanishing
after the spatial integrations.  This can be non-perturbatively
realized on the lattice by the periodicity of the link variables.
Alternatively, if we take the spatial sites $n^i$ (that gives $x^i=n^i
a$ with the lattice spacing $a$) from $-N_i$ to $+N_i$ for $i=x,y,z$,
these terms $\propto z, y$ trivially disappear.  The spatial volume is
thus $(2N_x+1)\cdot(2N_y+1)\cdot(2N_z+1)\cdot a^3$ in the present
simulation.  The corresponding momenta are discretized as
$p^i a=2\pi k^i/(2N_i+1)$.  We also note that we impose the
anti-periodic boundary condition to avoid the singularity at
$|\bp_A|=0$ in Eq.~\eqref{eq:ur} or in Eq.~\eqref{eq:phase}.  This
means that $k^i$ should take half integral values from $-N_i+1/2$ to
$+N_i+1/2$.  We emphasize here that, if the spatial volume is large
enough, it does not matter whether the boundary condition is periodic
or anti-periodic.  This is simply a prescription not to hit the
singularity at $|\bp_A|=0$.  One might think that one could insert a
small regulator in the denominator like the $i\epsilon$ prescription.
Indeed, one could adopt the periodic boundary condition and could take
an average of contributions from six neighbors, $p_A^x=\pm\epsilon$,
$p_A^y=\pm\epsilon$, $p_A^z=\pm\epsilon$ around $|\bp_A|=0$.  As seen
in momentum space, however, such an averaging procedure is effectively
identical to imposing the anti-periodic boundary condition.  In any
case we should carefully treat the singularity at $|\bp_A|=0$ because
this infrared singularity is responsible for the axial anomaly, which
is manifested in the representation with Berry's
curvature~\cite{Son:2012wh,Stephanov:2012ki,Gao:2012ix}.

\begin{figure}
 \includegraphics[width=\columnwidth]{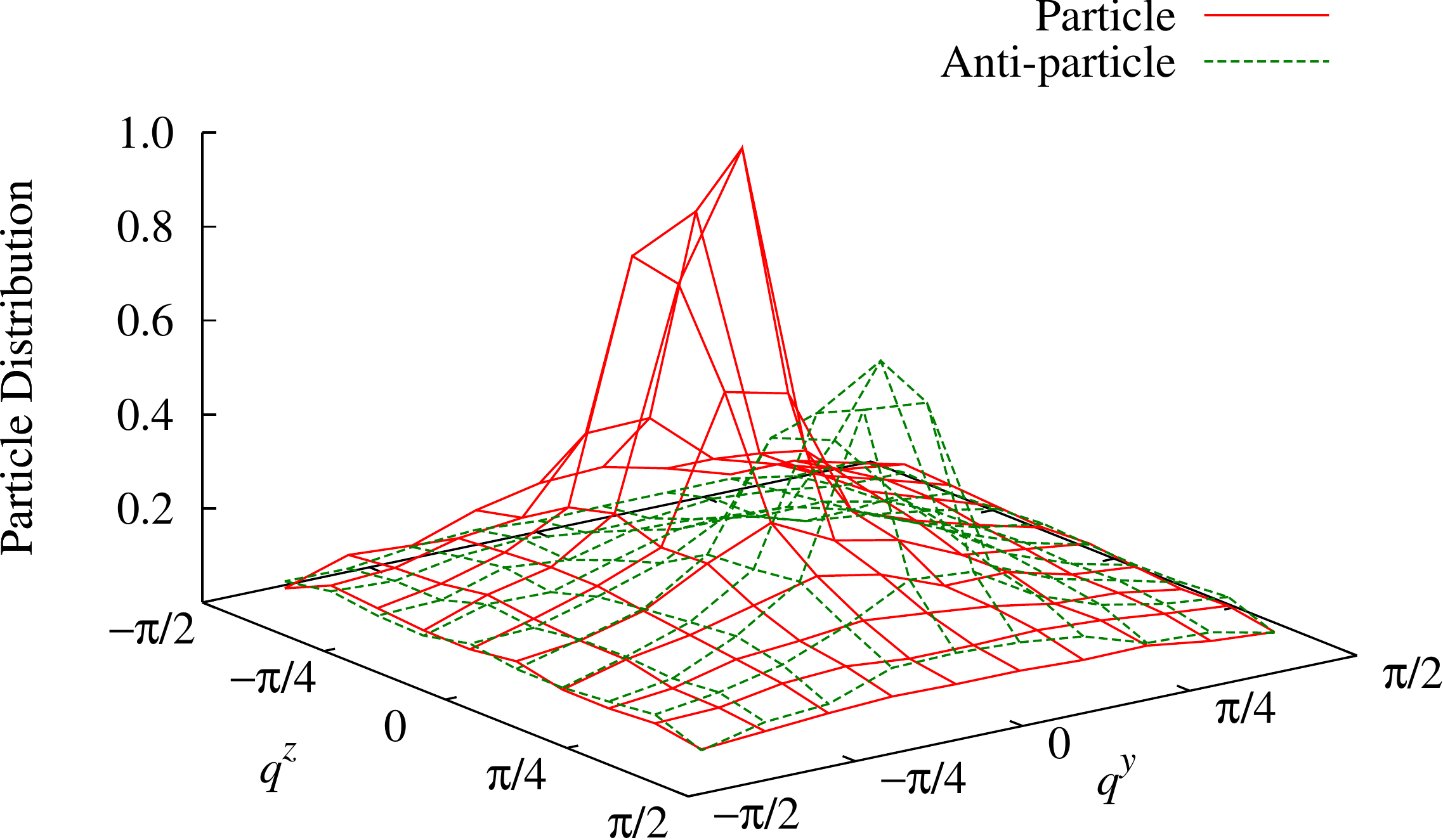}
 \caption{Distribution of the produced particles and anti-particles as
   functions of $q^y$ and $q^z$ (and integrated with respect to $q^x$;
   see the text).  The lattice size is $N_x=N_y=N_z=12$ and the
   magnetic fields are $B_\parallel=B_\perp=E_0/2$.}
 \label{fig:dist}
\end{figure}

If we integrate over all momenta, we cannot avoid picking up all
contributions from the doublers and the net particle production should
be then absolutely zero because the axial anomaly is exactly canceled
according to the Nielsen-Ninomiya theorem~\cite{Nielsen:1981hk}.  It
is thus crucial to get rid of the doubler contributions properly.  To
this end we limit our momentum integration range to the half Brillouin
zone only, i.e., $k^i$ from $-N_i/2+1/2$ to $+N_i/2-1/2$ (which is
used in Ref.~\cite{Gelis:2004jp} too).  We note that the axial anomaly
is correctly captured from the singular contributions in the
wave-function around $|\bp_A|=0$, though it usually appears near the
ultraviolet cutoff in a diagrammatic approach.  (We have numerically
checked that our results are robust against small shifts at the
momentum edges, which is also understandable from Fig.~\ref{fig:dist};
produced particles are centered with small momenta.)  One should not
be confused with the situation in Euclidean, i.e., static lattice
simulation in which the whole Dirac determinant should be evaluated to
identify the statistical weight for gauge configurations.  For the
present purpose, in contrast, we need the information on each momentum
mode, and so the computation is much more time consuming than
evaluating the determinant, but it is easier to get rid of the
doublers on the mode-by-mode basis.

%---   figure   ---%
%---   figure   ---%

When we change the lattice size and the lattice spacing $a$, we should
change a dimensionless $E_0 a^2$ accordingly to fix $E_0$ in the
physical unit.  In this work we choose a reference lattice size as
$N_x=N_y=N_z=8$ (i.e., $17^3$ lattice volume) and scale physical
quantities such as $E_0 a^2 \propto 17^2/(2N+1)^2$ with varying
$N_x=N_y=N_z=N$.  To simplify the notation let us absorb $e$ in the
field definition and omit $a$ hereafter.  Then, we take
$E_0(N=8)=0.1$ and change
$B_\parallel$ and $B_\perp$ from $0$ to $E_0$.  We set the duration as
$T=\sqrt{10/E_0}$ (that is, $T=10$ for $N=8$).  Our choice of $t_0$ is
$t_0=0.6T$ and we continue solving Eq.~\eqref{eq:Weyl_eq} up to
$t=1.2T$ with time spacing $\Delta t=0.02T$ using the 2nd-order
Runge-Kutta method (with which we carefully checked that the accuracy
is good enough for our present simulation).

We did not quantize the field strength unlike
Ref.~\cite{Bali:2011qj}.  Actually we compared results with
unquantized and quantized fields (in the unit of $2\pi/N_x N_y$ for
$B_z$ etc) and found that this would make only negligible difference.
Because our calculation is classical for gauge fields,
non-quantization of the field strength is harmless for our numerical
results, and moreover, technically speaking, we do not have to impose
a boundary condition for the gauge fields unless we solve the equation
of motion containing the derivatives of the gauge fields.

Even if $B_\parallel=B_\perp=0$, a finite $E_0$ would induce the
\textit{pair} production of particles and anti-particles under the
Schwinger mechanism (see Ref.~\cite{Dunne:2004nc} for a comprehensive
review).  Because we deal with Weyl fermion, the pair production is
always possible for any $E_0\neq 0$, which leads to a finite
distribution of particles with $-E_0 T \lesssim p^z\lesssim 0$ and
$p^x=p^y=0$ (and with $0\lesssim p^z\lesssim E_0 T$ for
anti-particles).  If $B_\parallel$ or $B_\perp$ is finite, this
distribution is smeared in momentum space as clearly observed in
Fig.~\ref{fig:dist} in which we present $\int(dq^x/2\pi) f(\bq)$ and
$\int(dq^x/2\pi) \bar{f}(\bq)$ as functions of $q^y$ and
$q^z$ in the case with $B_\parallel=B_\perp=E_0/2$.  In addition to
smearing, it is evident in Fig.~\ref{fig:dist} that the particle is
more enhanced over the anti-particle because of the $\calCP$-odd
background, which signals for the \textit{net} production of
right-handed particles.  (For the vector gauge theory the total
particle number is conserved as a result of summing right-handed
and left-handed contributions up.)

To close this section we shall make some remarks about the lattice
convention.  Taking the sum over dimensionless momenta $k^i$
corresponds to the phase-space integral $V\int d^3 \bp/(2\pi)^3$.  In
our definition $\beta_{\bq,\bp}$ has a mass dimension of $V$ and so
$f(\bp)$ has a mass dimension of $V$ too.  Then, $J^0$ and
$\boldsymbol{J}$ become dimensionless quantities.  In these final
expressions, however, the prescription to avoid the doubler
contributions makes the phase-space volume smaller by a factor
$N^3/(2N+1)^3$, which should be corrected in the end.  In this paper
we would not show the lattice version of expressions, but for example,
$q^i_{\pm A}$ in Eq.~\eqref{eq:currents} should be understood as
$\sin[(q^i\mp A^i)a]$.

%%%%%%%%%%   Numerical results   %%%%%%%%%%
\section{Numerical results}

\begin{figure}
 \includegraphics[width=\columnwidth]{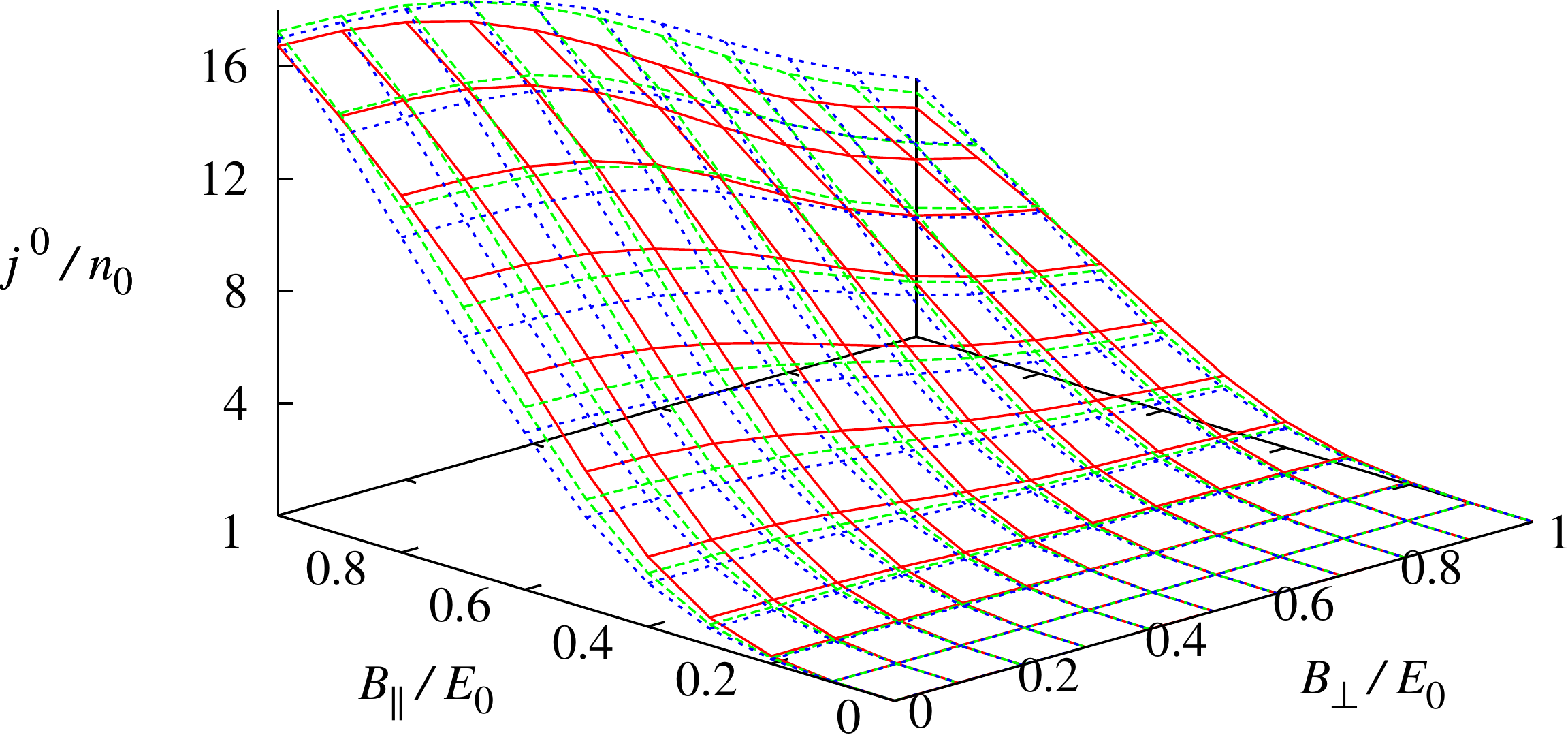}
 \caption{Net particle density $j^0$ normalized by $n_0$ in
 Eq.~\eqref{eq:n0} as a function of $B_\parallel$ and $B_\perp$.
 The lattice size is $N_x=N_y=N_z=8, 10, 12$ from the bottom (solid)
 to the top (dotted).}
 \label{fig:j0}
\end{figure}

In what follows we normalize $j^0=J^0/V$ and
$\boldsymbol{j}=\boldsymbol{J}/V$ by the following quantity:
\begin{equation}
 n_0 = \frac{E_0^2}{4\pi^3}\cdot T \;,
\label{eq:n0}
\end{equation}
which is the pair production rate $E_0^2/4\pi^3$ in the Schwinger
mechanism multiplied by the duration $T$, which gives us a natural
order-of-magnitude estimate.  Then, $j^0/n_0$ should remain to be a
reasonable number, which is confirmed in Fig.~\ref{fig:j0}.

%---   figure   ---%
%---   figure   ---$

Figure~\ref{fig:j0} clearly evidences for a non-zero value of the net
particle production for $\bE\cdot\bB=E_0 B_\parallel\neq0$.  We can
make it sure that the net particle production is vanishing when
$B_\parallel=0$.  This should be trivially so but a non-symmetric
treatment of coordinates would pick up unphysical artifact, which is
not the case in the present simulation.  In Fig.~\ref{fig:j0} the
lattice size is $N_x=N_y=N_z=8, 10, 12$ (i.e., $17^3$, $21^3$, $25^3$
lattice volumes) from the bottom (solid) to the top (dotted).  We can
conclude that $N_i$ dependence is rather mild, while it is still
there.  In fact $j^0$ shows moderate dependence on $B_\perp$, which
results from the lattice artifact.  In the steady system with constant
electromagnetic fields the anomalous particle production rate has only
weak dependence on $B_\perp$ and it slightly increases with increasing
$B_\perp$.  We can see a tendency that $j^0$ at smaller (larger)
$B_\perp$ is pushed down (up) as $N_i$ increases.

%---   figure   ---%
\begin{figure}
 \includegraphics[width=\columnwidth]{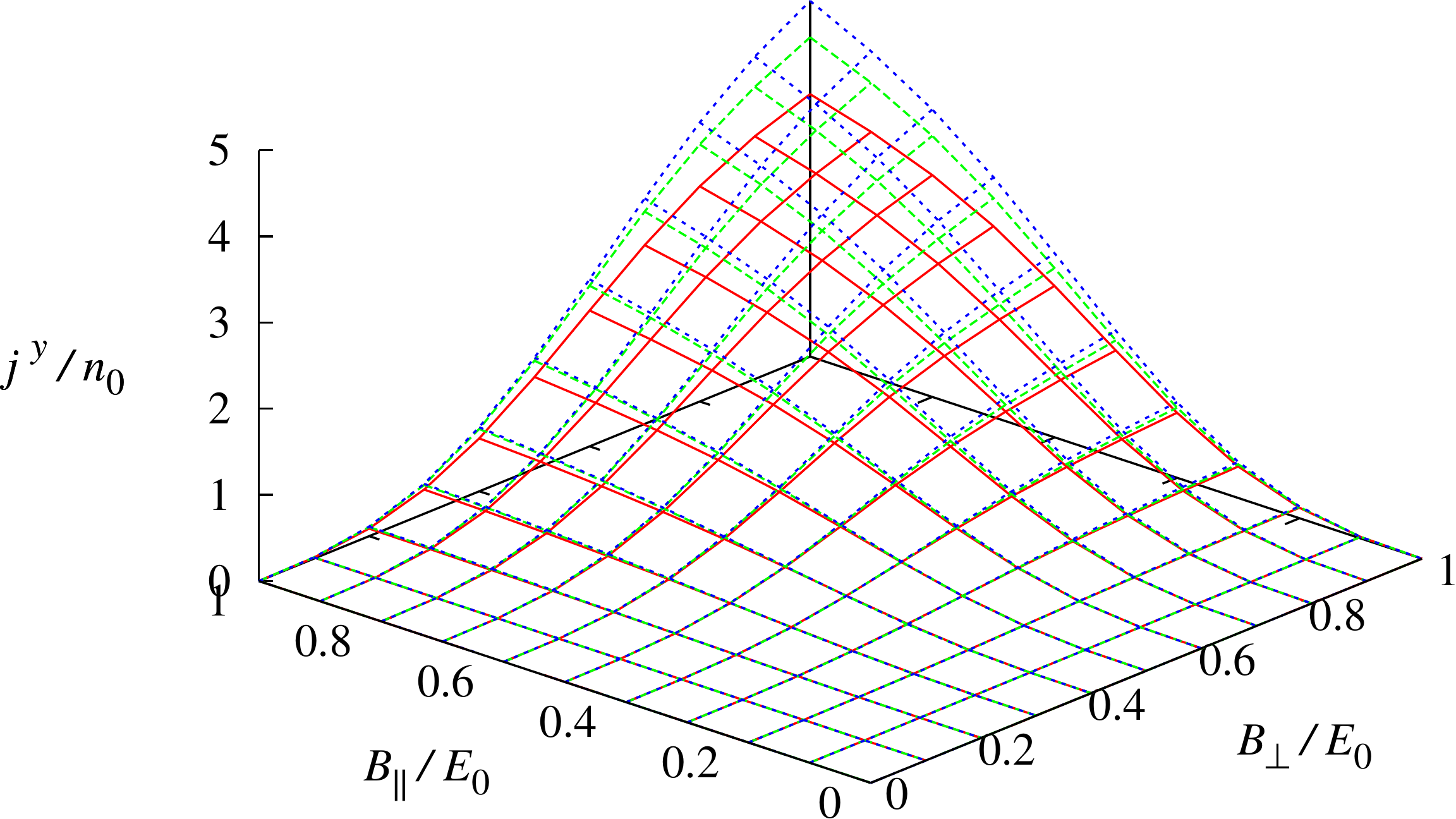}
 \caption{Net current density $j^y$ normalized by $n_0$ which is
   to be interpreted as the CME current.}
 \label{fig:jy}
\end{figure}
%---   figure   ---%

Now we are going to discuss $\boldsymbol{j}/n_0$ obtained in our
numerical calculation.  Here let us focus only on the chiral magnetic
current $j^y$ and postpone discussions about the anomalous Hall
current $j^x$ and Ohm's current $j^z$ to a separate
publication~\cite{pablo}.  It is interesting that our results in
Fig.~\ref{fig:jy} surely represents a non-zero CME current as
theoretically expected.  We see that $j^y=0$ when either $B_\parallel$
or $B_\perp$ is vanishing.  Because $j^y$ is the CME current, it is
naturally vanishing if $B_\perp=0$.  Furthermore, $B_\parallel=0$
means $\bE\cdot\bB=0$, and so, no \textit{net} particle production is
allowed (even though the \textit{pair} production is possible as long
as $\bE\neq0$).  The current is vanishing then due to the lack of
electric carrier.

%%%%%%%%%%   Lattice-size dependence and the response time   %%%%%%%%%%
\section{Lattice-size dependence and the response time}

%---   figure   ---%
\begin{figure}
 \includegraphics[width=\columnwidth]{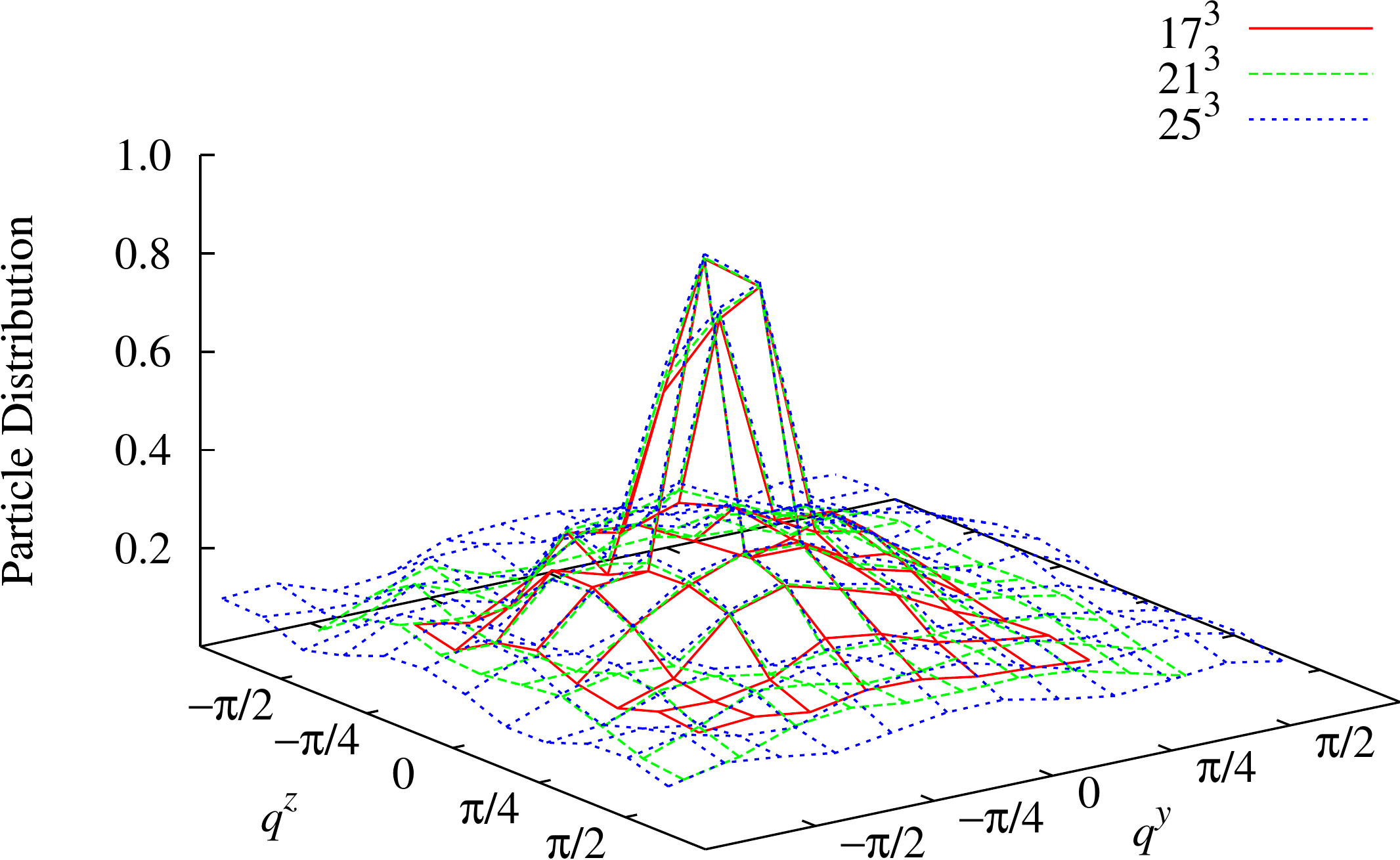}
 \caption{Distribution of the produced particles as functions of $q^y$
   and $q^z$ (and integrated with respect to $q^x$) for different
   lattice sizes $17^3$, $21^3$, and $25^3$.  The magnetic fields are
   $B_\parallel=B_\perp=E_0/2$ and the duration is $0.5T$.}
 \label{fig:dist2}
\end{figure}
%---   figure   ---%

The lattice-size dependence that we have seen is the ultraviolet
extrapolation, for which the box size is fixed and the lattice spacing
is decreased.  Therefore, $E_0$ in the unit of the lattice spacing was
rescaled to fix the observables in the physical unit.  In the present
case the numerical results should be stable for the ultraviolet
extrapolation.  We can understand this from Fig.~\ref{fig:dist2}.  As
mentioned before, the distribution function is well localized in
momentum space, and so the contributions from the high-momentum region
simply attach a tail in the particle distribution as shown in
Fig.~\ref{fig:dist2}.

In our definition $f(\bq)$ does not go to zero for $|\bq|\to \infty$
when $\bE\neq0$ and $\bB\neq0$.  If the subtraction of the zero-point
oscillation energy is normalized at $\bq|\to \infty$ we could remove
an irrelevant shift in the distribution function.  In principle,
however, such a constant shift does not affect $j^0$ and $\bj$:  the
particle and the anti-particle contributions cancel for $J^0$ and
there is no contribution to the angle integration of $\int d^3\bp$ for
$\bj$.  From this consideration we can expect better ultraviolet
convergence for $\bj$ because a constant shift vanishes for each of
the particle the anti-particle parts, while $j^0$ needs a delicate
cancellation.  We will see that this expectation is indeed the case in
our numerical simulation.

%---   figure   ---%
\begin{figure}
 \includegraphics[width=\columnwidth]{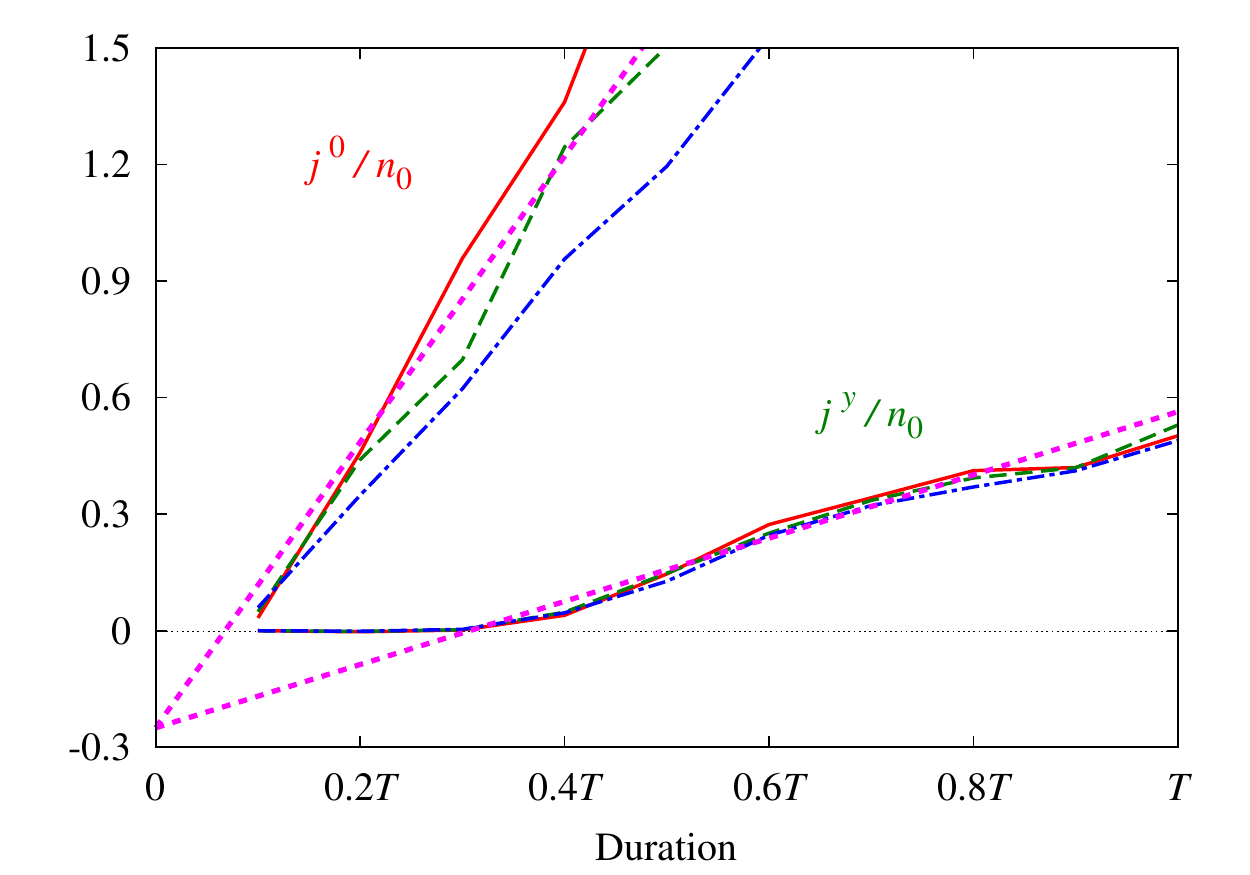}
 \caption{Profile of $j^0$ and $j^y$ as a function of the pulse
   duration normalized by $n_0$.  The lattice size is $N=8$ (solid
   line), $10$ (dashed line), $12$ (dash-dotted line).  The magnetic
   fields are chosen as $B_\parallel=B_\perp=E_0/2$.  The dotted line
   represents the analytically estimated rates for the particle
   production and the current generation (with a common offset
   $-0.25$).}
 \label{fig:time}
\end{figure}
%---   figure   ---%

It would be of pragmatic importance to consider $j^0$ and $j^y$ as a
function of the pulse duration, that is, the life time of the external
fields.  So far, we fixed it as $T=\sqrt{10/E_0}$ and let us vary it
now.  Theoretically speaking, for a sufficiently large duration, we
should expect linear dependence assuming constant rates of particle
production and current generation.  In contrast, for a small duration,
it is very difficult to predict what should happen \textit{a priori},
and this is why we must perform the numerical simulations.

Our numerical results in Fig.~\ref{fig:time} (for $N=8, 10, 12$
corresponding to $17^3, 21^3, 25^3$) where $B_\parallel=B_\perp=E_0/2$
is fixed clearly confirm linear dependence for large duration (apart
from small oscillation which is characteristic to magnetic phenomena).
Although $j^0$ shows some lattice-size dependence (due to the constant
shift in the ultraviolet region of the produced particle
distribution), two results for $j^y$ with different $N_i$'s are nearly
overlaid to each other.  This clearly indicates that, because the
constant shift drops off in $\int d^3\bq$, $j^y$ has no contamination
from the ultraviolet fluctuations.

We find for small duration a region of delay in which $j^0$ and $j^y$
remain almost vanishing.  The presence of such delay is intuitively
understandable: it should take some time for the wave-function to
transform from the free plane wave to the Landau-quantized one after a
sudden switch-on of the magnetic field.  We would emphasize that what
seems non-trivial in Fig.~\ref{fig:time} is that the delay for $j^y$
is more than three times larger than that for $j^0$.  In our
interpretation this difference in the onsets comes from the difference
in the rates.  In Fig.~\ref{fig:time} we showed the analytically
estimated rates (by the dotted line) using the formula in
Ref.~\cite{Fukushima:2010vw} with a common offset $-0.25$.  The onset
for $j^y$ is more delayed than that for $j^0$ since its rate
$\partial_t j^y$ is smaller.

Finally we briefly comment on the sensitivity when we change the box
size, i.e., the infrared extrapolation.  We kept the lattice spacing
(and so $E_0 a^2$) and changed the lattice size to find that there is
a subtle feature in the scaling behavior.  Since we need to go far
into technical details to address this problem, we will leave it for
another publication that will be focused on the lattice formulation
and the systematic checks of the lattice-size dependence.

%%%%%%%%%%   Summary and outlook   %%%%%%%%%%
\section{Summary and outlook}

We formulated the production of particles and anti-particles and
checked its validity by looking for a right-handed particle excess on
$\calCP$-odd backgrounds.  We limited our main concern to the CME
current at present, but this type of the calculation should be
applicable for a wider range of physical problems.  For example, the
(pair) particle creation at the horizon of the acoustic blackhole (see
Ref.~\cite{Balbinot:2014cfa} for a pedagogical article) attracts
attention, and it would be feasible to combine it with $\calCP$-odd
(or $\mathcal{T}$ irreversible) background effects, for which we can
apply our method.  Of course, as mentioned in the beginning, possible
applications should cover problems of sphaleron-like transitional
processes in the electroweak and the strong interactions on top of
arbitrary gauge fields including non-topological ones.

Here we emphasize the impact of the agreement between our numerical
results and the analytical expressions.  It is very non-trivial that
a direct evaluation of the current according to
Eq.~\eqref{eq:currents} could be consistent with the quantum anomaly.
Also we emphasize the importance of real-time character of the chiral
magnetic current generation.  The physical interpretation and the
theoretical estimate of $\mu_5$ are quite problematic and sometimes
misleading in the literature.  Our successful simulation is the first
step toward a realistic simulation in experimental setups without
$\mu_5$ in the system.

In this kind of approach to solve the equation of motion, the anomaly
arises from the infrared singularity around $|\bp_A|=0$.  The major
part of the discretization error might be attributed to
underestimating the singular contributions with coarse mesh.  This
implies that there may be a hybrid way to extract the singular terms
analytically and to calculate non-singular parts
numerically~\cite{pablo}.  This would reduce the lattice-size
dependence and we can make theoretical estimates with improved
reliability.  Though we have confirmed that our CME current is close
to the one in the continuum limit, it would be worth developing such a
hybrid algorithm.  Another direction for the improvement is to include
back-reaction from gauge fluctuations.  Once the back-reaction is
taken into consideration, it would capture the effect of the chiral
plasma instability~\cite{Akamatsu:2013pjd}.  These are works under
progress~\cite{pablo}.

\acknowledgments
This work was partially supported by JSPS KAKENHI Grant Number
15H03652 and 15K13479.

\end{document}